\documentclass[manuscript,screen,nonacm]{acmart}

\setcopyright{acmcopyright}
\copyrightyear{2023}
\acmYear{2023}
\acmDOI{XXXXXXX.XXXXXXX}

\acmConference[EuroUSEC'23]{European Symposium on Usable Security}{October 16--17,
    2023}{Copenhagen, Denmark}
\acmPrice{15.00}
\acmISBN{978-1-4503-XXXX-X/18/06}

\usepackage{multirow}%
\usepackage[title]{appendix}%
\usepackage{tikz}
\usepackage{textcomp}%
\usepackage{algorithm}%
\usepackage{algorithmicx}%
\usepackage{algpseudocode}%
\usepackage{listings}%
\usepackage{wrapfig}

\usepackage{float}
\usepackage{array,longtable}
\usepackage{enumitem}
\usepackage{tabularx}

\newlist{questions}{enumerate}{1}
\setlist[questions]{label*=RQ\arabic*}

\newcommand*\circled[1]{\tikz[baseline=(char.base)]{
            \node[shape=circle,draw,inner sep=1pt] (char) {\small #1};}}

\newcommand{\paragraphb}[1]{\noindent{\bf #1} }

\def \mwide {.25\textwidth}
\def \thin {.12\textwidth}
\def \cbox {$\bigcirc$} 
\newcommand{\st}[1]{
    #1 & \cbox{} & \cbox{} & \cbox{} & \cbox{} & \cbox{} & \cbox{} & \cbox{}
}

\begin{document}

\title{Introducing and Interfacing with Cybersecurity -- A Cards Approach}

\author{Ryan Shah, Manuel Maarek, Shenando Stals, Lynne Baillie, Sheung Chi Chan, Robert Stewart, Hans-Wolfgang Loidl and Olga Chatzifoti$^\dagger$}
\affiliation{%
    \institution{Heriot-Watt University, Edinburgh and Glasgow School of Art$^\dagger$}
    \country{UK}
}
\email{\{first.last\}@hw.ac.uk}

\renewcommand{\shortauthors}{Shah et al.}


%
%



\begin{abstract}
Cybersecurity is an important topic which is often viewed
as one that is inaccessible due to steep learning curves and a perceived requirement
of needing specialist knowledge. With a constantly changing threat landscape,
practical solutions such as best-practices are employed, but the number of critical cybersecurity-related
incidents remains high. To address these concerns, the National Cyber Security
Centre published a Cybersecurity Body of Knowledge (CyBOK) to provide a
comprehensive information base used to advise and underpin cybersecurity learning.
Unfortunately, CyBOK contains over 1000 pages of in-depth material and may not
be easy to navigate for novice individuals. Furthermore, it does not allow for
easy expression of  various cybersecurity scenarios that such individuals may be exposed to.
As a solution to these two issues, we propose the use of a playing cards format to provide
introductory cybersecurity knowledge that supports learning and discussion, using
CyBOK as the foundation for the technical content. 
Upon evaluation in two user studies, we found that 80\% of the
participants agreed the cards provided them with introductory knowledge of cybersecurity
topics, and 70\% agreed the cards provided an interface for discussing topics and enabled them to make links between
attacks, vulnerabilities and defences.
\end{abstract}



\maketitle

\section{Introduction}
\label{sec:introduction}

Cybersecurity remains a fundamental concern to users of computer
systems, with security often being overlooked due to its portrayal as a subject
pertaining to issues of perceived technical difficulty, steep learning curves and a requirement of specialist
knowledge and/or
expertise~\cite{harknett2009cybersecurity,hoffman2011thinking,asen2019you}. While the security foundations of
computer-based systems have improved over time, limiting the potential for, or mitigating the effects of, attacks
arising from  vulnerabilities, requires the involvement of all users of these systems (e.g. the general
population) and is a necessary step to improve the understanding of cybersecurity~\cite{adams1999users}. Moreover,
the increasing complexity and diversity of the threat landscape for
cybersecurity~\cite{wall2017crime,kaloudi2020ai,bahizad2020risks} further substantiates the need for improving
understanding of cybersecurity.
In the domain of software engineering, practical solutions to achieve this include activities such as the documentation
of vulnerabilities of computer systems and updating respective knowledge bases. Open databases such as the Common
Vulnerabilities and Exposures (CVE)~\cite{cve} and Common Weakness Enumeration (CWE)~\cite{martin2011cwe}, have played
a pivotal role in raising the awareness of known vulnerabilities such that appropriate defensive measures can be
developed or updated. While these reference databases are well maintained, they may still appear
complex to the general population and may contribute to the already existing problems of inaccessibility and
specialist requirements that are pinned against the topic of cybersecurity. Because of this, several knowledge bases
have been developed to inform and underpin cybersecurity education and
training~\cite{martinintroduction,newhouse2017national,ncscmsc}, which aim to address these issues at a high-school or
higher-education level. Although they may be a useful learning resource for providing key cybersecurity knowledge, their
primary purpose is to be used by those who are already knowledgeable in cybersecurity to develop further curricula to
teach those who may have little-to-no knowledge of cybersecurity. Furthermore, among these knowledge bases, there may be
some key topics which are not covered and their format and density may not be perceived as accessible to novice users.
Thus, this may directly impact one's ability to understand key cybersecurity topics but also to make links between these
topics to capture real-world cybersecurity scenarios. Ultimately, the weaknesses of existing solutions regarding limitations
of accessibility, steep learning curves and a perceived requirement of specialist knowledge/expertise, must be ameliorated
by a new solution that provides an answer to the following research questions. Specifically, can a new solution:

\begin{questions}[leftmargin=*,align=left]
    \item Provide introductory cybersecurity knowledge to novice users?
    \item Provide material for expressing interpretation and documentation of key cybersecurity topics, which can support independent learning and self-efficacy?
    \item Act as an index for the CyBOK knowledge base which provides an interface for discussion on key cybersecurity topics?
    \item Provide links between key cybersecurity topics, allowing the generation of concepts which can capture various cybersecurity scenarios?
\end{questions}

In this paper, we provide an answer to these research questions by proposing the use of a playing cards format as a medium to
provide: introductory knowledge of key cybersecurity topics, acting as an index for the CyBOK
knowledge base~\cite{cybokwebsite,martinintroduction}; support independent learning and self-efficacy; and allow for
links to be made between key cybersecurity topics to capture real-world scenarios. The novelty of this work is three-fold.
We first present the design principles for the cybersecurity cards to address these limitations.
Second, we provide an evaluation of the cards in a workshop with masters-level students to understand whether the cards
satisfy the aforementioned provisions. The output of this evaluation
is a second revised deck of the cybersecurity cards. Third, we carried out the same workshop but with a different
demographic to the first, with participants at late primary and early secondary school level (ages ranging from 10 to 15
years old, mean 12.8 years).

The remainder of this paper is organised as follows. Section~\ref{sec:background} provides background and related work,
as well as the selection procedure we applied to the production of our cybersecurity cards using the CyBOK knowledge
base and the limitations of other approaches. The design principles applied to the cybersecurity cards, as well as the
initial implementation (Version 1), are described in Section~\ref{sec:cards}. An evaluation of Version 1 of the cards
is provided in Section~\ref{sec:evaluation}. In Section~\ref{sec:cards1}, we present Version 2 of the cards as a result
of the findings from the first evaluation, as well as a further evaluation of the second version of the cards in Section~\ref{sec:evaluation2}. In
Section~\ref{sec:discussion}, we provide a discussion of the results from both evaluations and the paper concludes
in Section~\ref{sec:conclusion}.


\section{Background}
\label{sec:background}

The need for practical and easy-to-learn cybersecurity learning material is a constant problem which stems from
the evolving nature of cybersecurity and computing technologies as the number of connected
users and devices scales. In recent years, the number of critical cybersecurity
incidents have increased significantly, correlating with increasing numbers of
online users during the Covid-19 pandemic, for example, as well as an increase in the
adoption of various connected computer systems in day-to-day activities. Among these
incidents, research shows that around 95\% of cybersecurity breaches occur as a
result of human error~\cite{wef2020} and that organisations lack the sophistication, interest and/or
knowledge to handle these threats~\cite{ciscoreport2020,sophos2021}.

It has been shown that those in cybersecurity careers require a set of skills, involving the
abilities to carry out various tasks at any time in non-traditional environments, and adapt to
the dynamic nature of these environments~\cite{niccs2015}. In the domain of software engineering,
basic cybersecurity training such as password best-practices and multi-factor authentication are
employed for individuals to conform to, with the aim of alleviating concerns and mitigating the
potential for liabilities that arise as a result of cybersecurity-related
incidents~\cite{walters2011assessing,alotaibi2018review,sarginson2020securing}.
It has been identified that a large
number of Android applications contain security-related code snippets copied and pasted from Stack Overflow,
of which nearly 98\% contained at least one insecure code snippet~\cite{fischer2017stack}.
The value of security information depends strongly on its source~\cite{rader2015identifying} and reputable information
sources are only useful, so long as they are well-understood and perceived as
actionable~\cite{thomas2019educational,redmiles2020comprehensive}. While sites such as StackOverflow are reputable
for providing actionable solutions, it is clear that the security of solutions
are not well understood.
For novice individuals, such as those who write
and/or deploy software code without formal software engineering training, it may not be true that
they may fully comprehend the impact of not adhering to security best-practices.

To address this, various curricula guidelines and knowledge frameworks have been developed for
cybersecurity, covering a range of fundamental topics ranging from software and hardware security, to networks and
cyber-physical systems. The Joint Task Force (JTF) on Cybersecurity Education proposed a draft of
curricular guidance on cybersecurity to support educational efforts in the USA~\cite{bishop2017cybersecurity}.
They designed a framework model for a body of knowledge that covers six knowledge areas which
several concepts span over, targeting specific disciplines and application areas that pertain to
the demographic of cybersecurity professionals. The National Initiative for Cybersecurity Education
(NICE)~\cite{newhouse2017national} is a cybersecurity workforce framework, developed by NIST in the USA, which aims to provide
a foundation for describing and sharing information about knowledge, skills and abilities in
cybersecurity to strengthen an organisation's cybersecurity. The National Cyber Security Center (NCSC) in the UK proposed a Certified Master's
Program that defines several pathways to address knowledge and skill gaps in cybersecurity education,
which describe what topics must be covered and to what depth~\cite{ncscmsc}. While all these frameworks
tend to agree on key cybersecurity topics that must be understood, they only promote greater
emphasis on a subset of topics. For example, NICE covers a wide range of key topics but gaps exist
such as with topics related to cyber-physical systems and human factors. The NCSC Certified
Master's Program does not place much emphasis on attacks and defences, but in contrast focuses on key topics such
as software security.

\begin{figure*}[t]
    \centering
    \includegraphics[width=\linewidth]{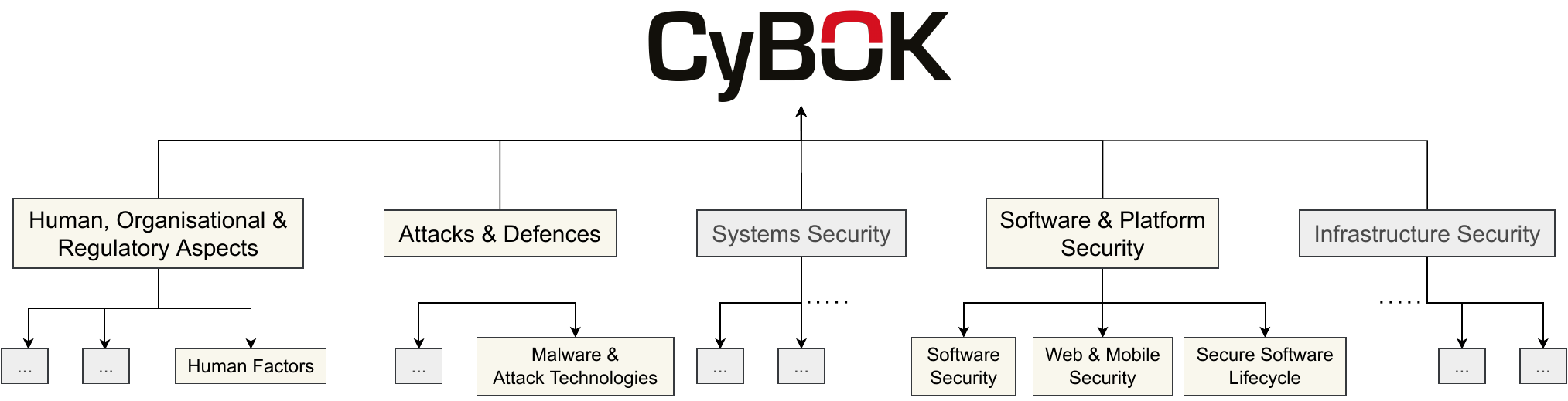}
    \caption{
        \centering Partial View of CyBOK 1.1 Knowledge Tree~\cite{cybokwebmobile}. The knowledge areas and topics that are highlighted show the subset of the CyBOK knowledge base
        that was selected due to the link to the domain of software engineering.
    }
    \label{fig:cybokfulltree}
\end{figure*}

The Cybersecurity Body of Knowledge (CyBOK) is a knowledge base developed by University of Bristol funded by the
NCSC. It was developed to encompass the wide variety of topics within the field of cybersecurity and to show
that it also spans multiple disciplines. In practice, it has been successful in providing a framework for NCSC certified
degrees and academic/professional training programmes~\cite{cybokcases}. CyBOK is decomposed into 21 knowledge areas (KAs) (as of version 1.1), each introduced by a
reference document and a set of topics presented as a branch of the overall {\em Knowledge Tree}
(Figure~\ref{fig:cybokfulltree})~\cite{cybokwebsite}. Each of these knowledge areas are organised into a hierarchy
of between 3 to 5 categories that present as a tree of topics.
For each KA in CyBOK, there are a number of chapters that form an encyclopedic collection of
knowledge of key concepts that are based on state-of-the-art academic literature.
These key concepts are known as {\em Topics}, with some {\em Topics}
decomposed further into a set of more specialised subjects ({\em Sub-Topics}). For example, the category of
{\em Software Security} in the {\em Software and Platform Security} KA contains 4 overarching themes, split into 20 sub-topics
(e.g. structured output generation vulnerabilities), each of which describe further specialised information (e.g. sql injection).


It has been shown that, in comparison with other knowledge frameworks, CyBOK covers a wider range
of knowledge areas and does not have gaps that are present within other frameworks~\cite{hallett2018mirror}.
While CyBOK facilitates a body of knowledge which attributes to the production of material for cybersecurity
education and professional training, there are some weaknesses which may render it an inaccessible resource to
more novice individuals such as those in the domain of software engineering who write or deploy code with no formal
software engineering training. First, the links between meaning and relationships among topics and sub-topics
vary across the entire Knowledge Tree, which prevents easy expression of various cybersecurity
scenarios. Second, the material across the CyBOK knowledge base and its indexing structure is not easy to traverse
for novice users. Gonzalez et al.~\cite{gonzalez2022exploring} show that it would be difficult for novice individuals
to infer the links between various topics, given that some follow either a single predominant theme or span several
topics themselves. Ultimately, to support novice users as well as those more experienced, key cybersecurity knowledge
provided by knowledge bases such as CyBOK require adequate presentation that can facilitate independent learning whilst
also providing a suitable interface for discussion of various cybersecurity scenarios to make the links between meaning
and relationships among topics.

Aside from knowledge frameworks,
cybersecurity information
has also been presented in other ways.
Capture the Flag (CTF) activities provide a series of competitive exercises used to find
vulnerabilities in computer systems and applications and have shown to be a valuable
learning tool~\cite{trickel2017shell,cybokctf,swann2021open}. Thomas et
al.~\cite{thomas2019educational} propose the use of a collectible
card game (CCG) as a means of teaching cybersecurity to high school students, given the benefits of
prevalence culturally to all age groups (familiarity) and encouraging the understanding
of competitive strategy and mistake-making as a way of learning~\cite{turkay2012collectible}.
Anvik et al.~\cite{anvik2019program} propose the use of a web-based card game for learning
programming and cybersecurity concepts, using simple vocabulary to create ubiquitous learning
experiences. Denning et al.~\cite{denning2013control} propose the use of a tabletop card game,
Control-Alt-Hack, with the aim of providing awareness training for cybersecurity,
arguing that playing card games can provide a reachable foundation for providing digestible
cybersecurity information to large audiences. However, while these gamified approaches show
various levels of success, there are limitations. First, many of these different approaches have
different target personas and goals. Second, card game approaches such as
Control-Alt-Hack~\cite{denning2013control} do not cover a broad range of key cybersecurity topics,
such as those identified by knowledge frameworks such as CyBOK, and do not adequately highlight
the links between vulnerabilities, attacks and defences. Specifically, attacks are typically
highlighted first, which does not help users understand how attacks present themselves (opportunistic
vulnerability targeting) and how to protect against them. Third, while CTF activities, for example, are
beneficial in this aspect~\cite{swann2021open,trickel2017shell}, a key disadvantage pertains to
novice users wherein competitions rely on technical expertise and the ability to traverse computer systems
using various command-line tools and other bespoke applications~\cite{ford2017capture}, or requiring (at
a minimum) a basic understanding of cybersecurity concepts in order to progress in finding vulnerabilities~\cite{mcdaniel2016capture}.

\section{Cybersecurity Cards -- Version 1}
\label{sec:cards}

To answer the research questions presented in Section~\ref{sec:introduction}, we propose a novel approach to represent key cybersecurity
knowledge, utilising specially designed playing cards. In recent years, it has
been shown that specially designed cards used as a tool for education attributes to positive outcomes in the space of
learning, attitudes and critical thinking skills~\cite{kordaki2016computer,kordaki2017digital}. With a large portion
of the general population familiar with card-based games, it is clear that a playing card format is a good medium,
both physically and digitally, with relatively intuitive logic if they are designed well. The cards proposed in this work
were developed by a team of experts in the fields of cybersecurity, human-computer interaction and games experts. A
subset of the cards we propose in this section have been made available for viewing and can be found
online\footnotemark[1]. The full set of cards will be made
publicly available under the CC BY-NC-SA Creative Commons license after the research project ends in January 2024.

\footnotetext[1]{\url{https://anonymous.4open.science/r/cybersecurity_cards-9F00/} (anonymous repository for double-blind reviewing purposes)}

\subsection{Focus of Knowledge}

In order to establish a good grounding for an evaluation of a card-based solution to address the proposed research
questions, an important factor is to determine what information the cards represent ($RQ1$). In this work, we leverage
the information found in the CyBOK knowledge base and produce the cybersecurity cards to address the limitations of
the knowledge base. Specifically, the focus of knowledge we chose to take is in the domain of software engineering
and secure coding, given recent increases in threats pertaining to this domain~\cite{saxena2020impact,pranggono2021covid} and the
observation of finding people with skills in secure coding reported as the most difficult task~\cite{willetts2014cyber}.
The topics that fall within the scope of the focus of knowledge in this work were derived from a filtering and
selection procedure, where the CyBOK knowledge tree was traversed from Knowledge Areas to Topics. Given that not all
KAs in CyBOK are related to the domain of software engineering, we selected topics that relate to this theme.
As highlighted in Figure~\ref{fig:cybokfulltree}, we chose the following knowledge trees to traverse:
{\em Human Factors}, {\em Malware \& Attack Technologies}, {\em Software Security}, {\em Web \& Mobile Security} and the
{\em Secure Software Lifecycle}. For each KA, we follow each branch of the tree to the leaf nodes. For each of the
leaves, we pick up the topic and provide a better description of that name to create the first version of the card. If
there are duplicates or highly similar topics in the CyBOK mapping, these are either merged together or split into
distinct types. For example, CyBOK's information on {\em Cross-Site Request Forgery} is mapped to two vulnerabilities:
{\em Inadequate Data Authenticity and Origin Verification} and {\em Inadequate Session Expiration}. Any leaves that fall
within Topics that, for example, may not follow a single predominant theme and do not relate to the selected focus of
knowledge are cut out. The category of {\em Software \& Platform Security} is directly related to the scope of this work
whereas others are not, such as {\em Forensics} which involves identifying and analysing data to support legal
proceedings. This process was carried out by two academic researchers in cybersecurity who decided on the final first
set of cards in consensus.

\begin{figure}[h]
    \centering
    \includegraphics[width=0.5\linewidth]{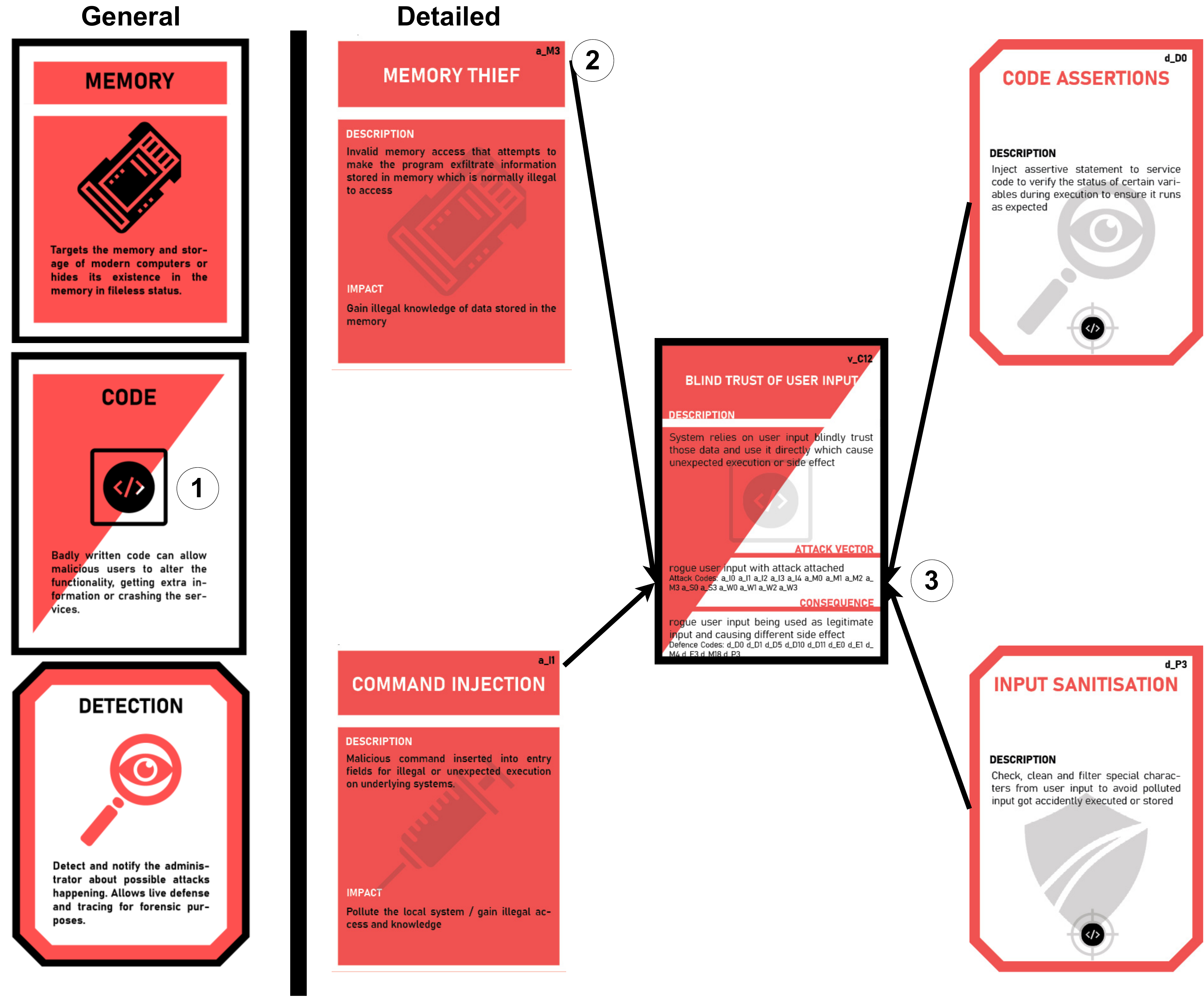}
    \caption{\centering Organisation and Design of Cybersecurity Cards (Version 1). General cards contain the icon of its corresponding class (1), with Detailed cards containing an identifier code (2) which is used to make links between attacks, defences and vulnerabilities (3). The arrows demonstrate the links between the presented attacks, vulnerabilities and defences.}
    \label{fig:carddesign}
\end{figure}

\subsection{Design and Structure}

The playing cards format is easy to handle in both a physical or digital medium~\cite{altice2014playing} and are
designed to be handled individually or combined into a set of cards (known as a {\em deck}). The deck of cards is first
split into three distinct classes: \textbf{Vulnerability}, \textbf{Attack} and \textbf{Defence}, following suit from
common threat modelling approaches and security catalogues that encompass the vulnerability-attack-defence
trichotomy~\cite{kordy2014attack,ferro2021human,tatam2021review,cve,martin2011cwe}. Within each of these classes, the
deck is further split into {\em \textbf{General}} and {\em \textbf{Detailed}} cards (Figure~\ref{fig:carddesign}).
General cards represent a type of one of the classes (e.g. {\em Code}), while Detailed cards provide lower-level,
concrete examples that relate to the General card.
In terms of design (Figure~\ref{fig:carddesign}), attack and vulnerability cards are encased in a square border, with
attack cards filled with a solid red colour while vulnerability cards are signalled with a diagonal line to separate
white from red. The defence cards resemble an octogonal shape (akin to a shield) which has a white fill colour. The
diagonal separation of white and red in vulnerability cards aims to highlight vulnerabilities as the central focal point.
General cards in the deck include: a title representing the KA topic; a type represented
by a symbol (see \circled{1}); and a description related to the topic it represents. Detailed cards are assigned a
unique identifier in the form {\em $a\_Bi$} in the top right corner \circled{2}, where $a$ refers to the class of the
Detailed card, $B$ referring to the first letter of the General card it is categorised under and $i$ acting as an
index number in the set of $N$ Detailed cards for the category $B$. The symbol for the Detailed card's category is
made opaque in the center of the card, behind the description of the Detailed card. Attack cards also contain a
description of the impact of the attack. Defence cards contain a target symbol to help further identify the
vulnerability it aims to protect against attacks. Vulnerability cards also describe an attack vector with associated
attack card identifiers, as well as the consequence of the vulnerability with associated defence card identifiers
\circled{3}. The aim of the identifiers is to provide a means to make links between key cybersecurity topics by allowing
users to capture various attack-defence scenarios that revolve around specific vulnerabilities ($RQ4$). In total, there are 124 cards in the deck which is composed of 30 vulnerability cards, 32 attack cards and
47 defence cards, each of which are categorised under one of the 15 General cards.

\subsubsection{Relationships Between Cards}
Given that cybersecurity stems from conflict between attackers and defenders targeting one or more vulnerabilities,
creating a capturable {\em many-to-many relationship} is essential when introducing cybersecurity concepts. Thus, the
cybersecurity cards should represent this relationship, where multiple attacks can target multiple vulnerabilities that
can, in turn, be mitigated or countered by multiple defences. While CyBOK does contain information about these
relationships, it is hard to infer these from implicit references within the reference material. Thus, the cards aim to
act as an index for the CyBOK knowledgebase by providing a means for presenting these implicit relationships in a manner
that supports independent learning and self-efficacy ($RQ2,RQ3$).
In Figure~\ref{fig:carddesign}, we can see that the vulnerability card {\em "Blind Trust of User Input"} is linked to a
set of identifier codes \circled{3} for a number of related attacks and defences. This example shows a link to the
        {\em "Command Injection"} and {\em "Memory Thief"} attacks ($RQ4$). The first attack involves the execution of unauthorised
commands to a system (which may be input by a user) and the second involves the stealing of confidential information
from memory. In contrast, two defence examples are shown as links to the attacks and the vulnerability, which include
        {\em "Code Assertions"} and {\em "Input Sanitisation"} which involve monitoring code execution and sanitising user
input to eliminate malicious code or escape characters, respectively.



\section{Evaluation}
\label{sec:evaluation}

The logical next step after designing the cybersecurity cards was to conduct a user
evaluation to see if the participants could use the cards, and whether the cards
provided a clear communication of these topics and an interface for discussing them.

\subsection{Methodology}

Participants were recruited to take part in a workshop, advertised to university students, where they interacted with
the cybersecurity cards in order to analyse and devise a cybersecurity scenario. Ethical approval was granted by the
University ethics committee for the workshop recruitment and procedure before the study took place. A participant
information sheet was provided via a link in the online registration form, as well as being used to obtain written
consent from participants. Furthermore, the workshop was held in Spring 2022 and we had a special clause added to state
we would follow local government guidelines on the Covid-19 protocol~\cite{covidspring22} which provided guidance on
reducing risks from transmitting Covid-19, such as staying home if participants had symptoms or tested positive for
Covid-19. In total, we managed to recruit 13 participants but only 11 had usable data. Upon registration for the
workshop, we found that the 11 participants were masters-level students who were between 22 and 35 years old
(mean 26.3 years). The participants were enrolled in a conversion masters program in computer science and came from
either an engineering (6), mathematics (3), computing (1) or biology (1) background. When asked if the participants had
any prior experience or skills with cybersecurity, only one participant mentioned they had experienced anomaly detection
but others stated they had no prior experience or skills in cybersecurity. Further, when asked if they had any
experience or skills with secure coding, 8 of the participants said they had no experience or skills with secure coding,
2 responded neutral and 1 said they had some experience. We hypothesise that the low number of participants to be
attributed to the Covid-19 global pandemic leaving individuals with more dynamic priorities and commitments.

\begin{table}
    \centering
    \begin{tabularx}{\linewidth}{l|Xl@{}r}

        \textbf{Phase} & \textbf{Activity} & \textbf{Rationale} & \textbf{Duration (min)} \\

        \midrule

        Prep & Participant information sheet & Inform participants & 5 \\
        & Informed consent form & Inform consent & 5 \\
        & Demographic Questionnaire & Participant profiles & 5 \\

        \midrule
        Activity & Introduction to workshop theme & Introduce theme & 10 \\
        & Introduction to cybersecurity cards and experts roaming around the room providing help when needed & Familiarity with cards & 10 \\
        & Theme exploration -- devising attack and defence scenarios & Cybersecurity trichotomy & 20 \\

        \midrule
        Post & Cybersecurity Questionnaire & Evaluate cards & 10 \\

        \midrule
    \end{tabularx}
    \caption{Workshop Structure Showing Phases, Activities and Duration (min)}
    \label{table:workshopstructure}
\end{table}


An overview of the workshop structure can be seen in Table~\ref{table:workshopstructure}. Participants were given the
information sheet, consent form and demographic questionnaire to fill out in advance of the workshop, with copies of the
consent form brought on the day as back ups in case they forget to send or bring the consent form. When they arrived
they were split into three groups (consisting of three or four participants).
During the workshop, participants were first introduced to the theme of the workshop which was {\em Code Security} --
the practice of developing software that embeds security and best-practices into the code. After this, they were given
an introduction to the deck of cybersecurity cards where they had time to
explore and familiarise themselves with the various attacks, defences and vulnerabilities. Cybersecurity experts that
were present would roam around the room and were able to answer any questions participants may have had when going
through the deck. After this, they would use the cybersecurity cards to find vulnerabilities and devise attack and
defence scenarios within the theme of the workshop. At the end of the workshop, each
participant was individually asked to fill in a self-assessment questionnaire online (Appendix~\ref{app:questionnaire})
on how using the cards contributed to the themes described in Section~\ref{sec:results1}.
The rationale behind this questionnaire was to collect both quantitative and
qualitative data regarding a participant's experience of interacting with the cybersecurity cards. The quantitative
aspect of the questionnaire (Q1) makes use of a 7-point Likert scale, ranging from strongly disagree to strongly agree,
to assess the performance and effort from participants while using the cards. The qualitative aspect (Q2--5) aims to
understand motivations and thoughts behind using the cards. Question 2 and 4 gave participants a series of options
(checkboxes), which were then coupled with a follow-up question (Q3 and Q5 respectively) to further elaborate on their
choices. Two experienced postdoctoral researchers independently analysed and coded the qualitative responses, grouping
them into themes, which were then discussed systematically and final agreements on codes were made in consensus.
This analytical technique has been applied successfully in various bodies of
work~\cite{zade2018conceptualizing,chinh2019ways}. The data from questionnaires is presented as italics in quotation
marks, alongside a participant ID (e.g. W1P1 for this first workshop) where relevant. 

\subsection{Results}
\label{sec:results1}

\begin{figure}[h]
    \centering
    \includegraphics[width=0.65\linewidth]{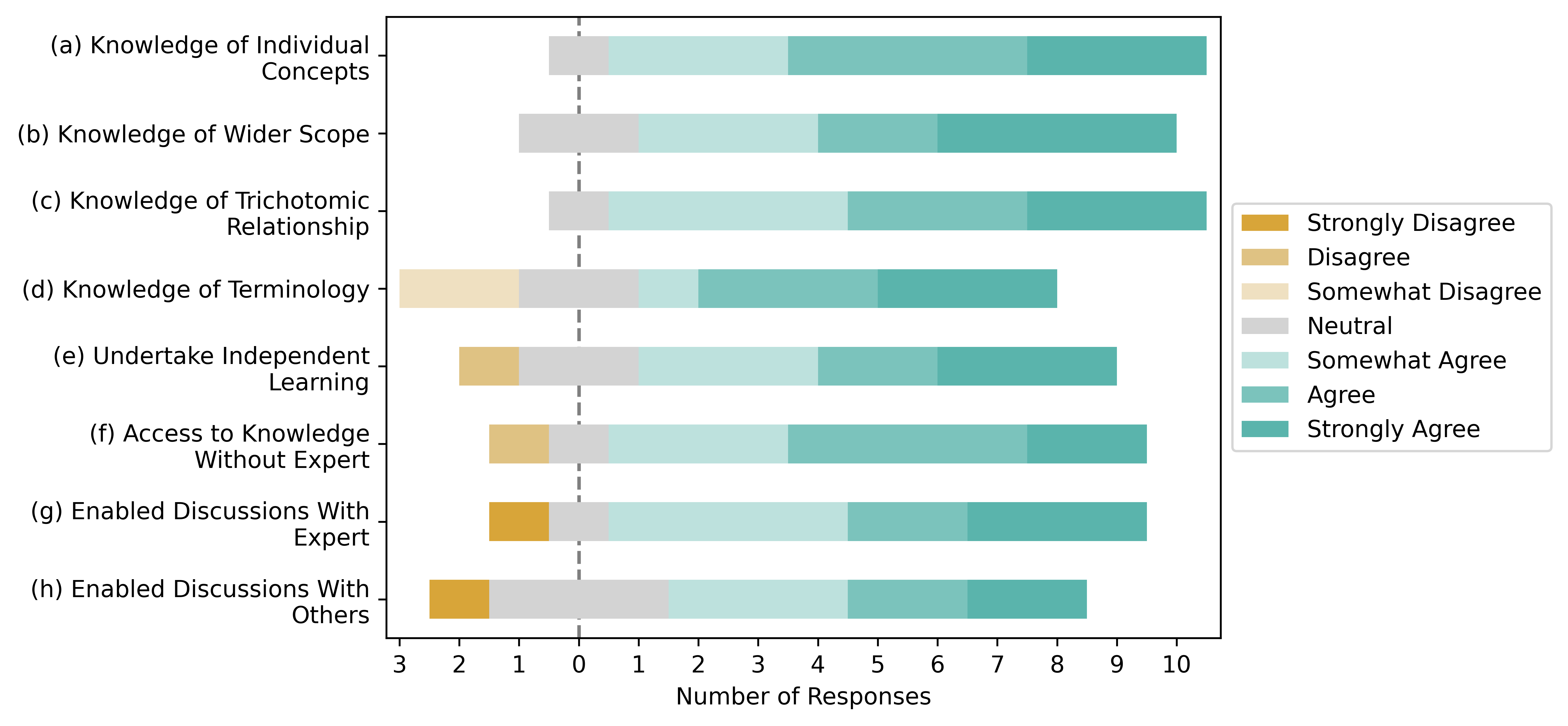}
    \caption{\centering Results of Workshop 1. This figure shows an overview of the responses for Q1 in Appendix~\ref{app:questionnaire}. (1) Providing a knowledgebase = (a)--(d); (2) Independent learning and self-efficacy = (e)--(f); (3) Interface for discussion: (g)--(h).}
    \label{fig:ws1likert}
\end{figure}

An overview of the results from this evaluation can be seen in Figure~\ref{fig:ws1likert}. For subsequent discussion,
the results from the questionnaires will initially follow three themes, noted by the title of each subsection.

\subsubsection{Providing a Knowledge Base}
\label{sec:provideknowledge1}

The first theme aimed at determining whether our cybersecurity cards provided users with introductory cybersecurity
knowledge, whilst also supporting learning in a well-documented manner. Providing a knowledge base looked into four
aspects: providing knowledge of individual concepts, cybersecurity terminology, wider scope of topics, and the
trichotomy of attacks, defences and vulnerabilities (\textbf{a--d}) in Question 1 of the questionnaire in
Appendix~\ref{app:questionnaire}). Overall, we observed that our cards were able to effectively provide
key cybersecurity knowledge, adapted from the CyBOK knowledge base, in an accessible manner. Interestingly, with regard
to individual concepts and the relationships between attacks, defences and vulnerabilities, three participants strongly
agreed with this. Furthermore, four participants also strongly agreed that the cards were able to provide them with
knowledge about wider scope. For cybersecurity terminology, only two responded with somewhat disagree, which we believe
may be due to the text on cards using too much technical terminology. It is clear that the cards have indeed achieved a
positive result regarding providing users with knowledge of fundamental cybersecurity knowledge ($RQ1$), as well as
providing links between topics that allows the generation of concepts which capture the relationships between attacks, vulnerabilities
and defences in scenarios pertaining to the theme of {\em Code Security} ($RQ4$).

\subsubsection{Independent Learning and Self-Efficacy}
\label{sec:independentlearning1}

With regard to independent learning and self-efficacy (\textbf{e,f}), we found that 8 participants agreed the cards enabled them to
undertake independent learning with three strongly agreeing. Furthermore, we found that 9 participants agreed the cards
provided them with access to key cybersecurity knowledge when one of the rotating experts was not present in the group.
In both cases, only a single participant disagreed. Ultimately, the results demonstrate that our cards approach provides
a means for expressing interpretation and documentation of key topics that supports independent learning and self-efficacy ($RQ2$).

\subsubsection{Providing an Interface for Discussion}
\label{sec:provideinterface1}

The final theme looks at whether the cards provide users with an interface for discussing key cybersecurity topics (\textbf{g,h}).
Interestingly, we found that more participants agreed that they could hold discussions on these topics with the expert
(9), in comparison with other participants in their group (7). However, while both cases show a majority in agreement,
when discussing topics with others in their group more participants remained neutral as opposed to disagreeing. While
we see a positive outcome regarding the cards providing an interface for the discussion of key cybersecurity topics ($RQ3$), it
is important to determine what was not clear and why this was the case, as well as understanding how the cards
could be improved.

\subsubsection{Understanding Drawbacks Of Using The Cards}
\label{sec:limitations1}

The next part of the questionnaire involves questions that were designed to help uncover and understand any limitations
of the cybersecurity cards, such that improvements can be made to better fulfill the goals of the research questions
described in Section~\ref{sec:cards}. The first step was to determine which category or subset of the cybersecurity
cards the participants did not use and why this was the case. We found that the least used cards were the general and
detailed defence and vulnerability cards. The next question asked participants why the cards they had selected were not
used, one participant stated that {\em "general cards gave some idea about the content"} (W1P6) and another stated that
they were only {\em "engaged in attack and defence"} (W1P11). With this said, however, most participants described that
they had used all of the categories, for example stating they had used {\em "at least one card [from] each"} (W1P1) or
only {\em "concentrate[d] on a few cards"} (W1P9). This suggests that while the cards which were said to not have been
used may not have been critical to discussions they had about certain topics, they may have still been looked at and
thought about. We then asked participants about how the cards could be improved. We found that 5 participants recorded that
the number of cards in the deck was too high. This may be due to the use of a physical medium and the number of
categories and types of each General card and is further suggested by 3 other participants stating that there were too
many types or categories. Interestingly, 5 participants stated that both the color coding of the types/categories was
not clear, as well as the relationships between the cards, with one participant describing that {\em "Threat cards
[were] difficult to handle/understand"} (W1P7). This may also correspond with the feeling that the number of cards was
too high, but also a perceived difficulty of understanding links between the cards and the terminology used. In contrast
to this, however, one participant mentioned that the {\em "Cards provides an entry point for more detailed scenarios of
cybersecurity and helps to create relationship between attack and defense situations more clear"} (W1P4). Interestingly,
3 participants recorded that the information on the cards were too abstract, with no participants stating that they were
too detailed. Thus, this helps reinforce the understanding that the difficulty relating to understanding the cards may
be linked to a lack of clarity on the relationships between them.



\section{Cybersecurity Cards -- Version 2}
\label{sec:cards1}

Upon review of the results from the evaluation questionnaire, the
cybersecurity cards were in places redesigned and the new deck is hereafter
referred to as {\em Version 2}. As well as providing the subset of the Version 1 cards for viewing, the equivalent
subset for Version 2 of the cards have also been made available and can be found
online\footnotemark[2]. The full set of cards will be made publicly available under the CC BY-NC-SA Creative Commons license after the research project ends in January 2024.

\footnotetext[2]{\url{https://anonymous.4open.science/r/cybersecurity_cards-9F00/} (anonymous repository for double-blind reviewing purposes)}


\subsection{Structural Redesign}

\begin{figure}
    \centering
    \includegraphics[width=0.75\linewidth]{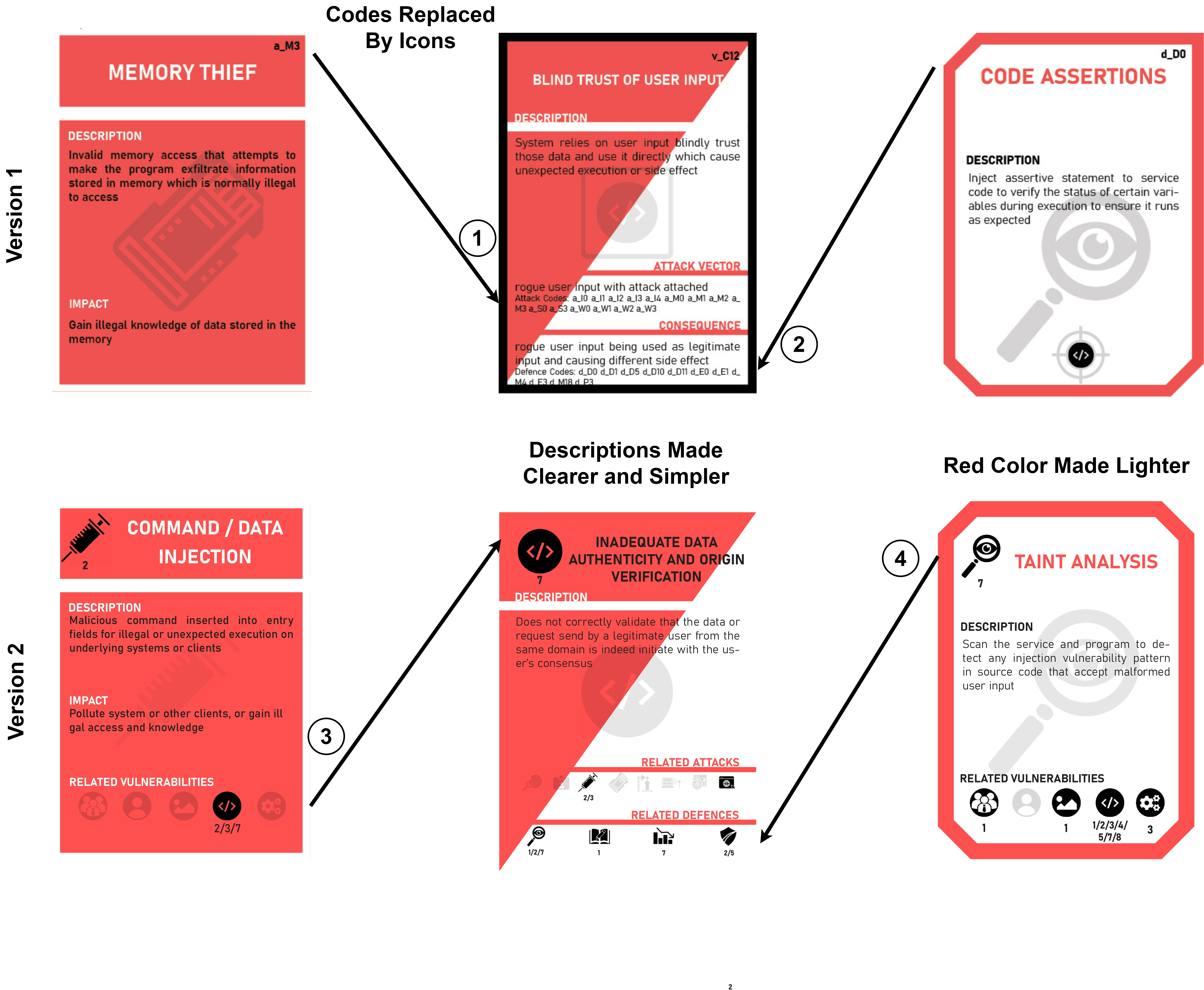}
    \caption{\centering Design Comparison Between Version 1 and 2. Version 1 identifier codes at the top right of each card (1) has been replaced by icons corresponding to its class and a unique class index number for each card (3). Links between cards using codes (2) have been replaced with the corresponding icons in vulnerabilities (4), and links between presented cards are demonstrated using the arrows.}
    \label{fig:v1v2comparison}
\end{figure}

One of the limitations that required addressing was with regard to
the number of cards in the deck, which some participants felt was
too high and {\em "difficult to handle and understand (W1P7)"}.
In contrast with a standard
deck of playing cards used for the likes of Solitaire and Poker, which has 52 cards,
Version 1 of our cards has more than double this amount (109) resulting in it being
perceived by participants as hard to handle. Thus, reducing the number of cards will likely be better received,
whilst still meeting our requirements.
In Version 2, the first major revision is reducing the number of cards in the deck. The first decision made with regard
to this was the removal of General cards (types of attack, defence or vulnerability), given that participants specified
these as the least used cards in the deck. In Version 2, General cards are replaced with a {\em glossary} (Appendix~\ref{app:glossary}) that
provides a description for each type, alongside a symbol associated with the type in the top left of the
card to help improve readability and visibility (\circled{4} in Figure~\ref{fig:v1v2comparison}). We also merged some
cards when distinctions were too specific, by identifying cards of overlapping topics and a set of 2 or more attack similar
attack cards were merged by a more general card. For example, the {\em Smudge Attack}, {\em Shoulder Surfing} and {\em Social Engineering} cards were merged to {\em Social Engineering}. This reduces the number of cards
in Version 2 to a total of 70 cards, made up of 20 attack, 20 vulnerability and 30 defence cards. By providing a
glossary, users can refer to this sheet as a form of guidance when looking at devising particular cybersecurity
scenarios.
Furthermore, we also added two
new classes of vulnerability. First, {\em Human} is split into
{\em User} and {\em Management}. The {\em User} class captures the impact of vulnerabilities from careless or malicious
users (e.g. insider attackers), while {\em Management} captures bad management practices such as poorly implemented
security policies. Second, we provide a {\em System} vulnerability type which
relates to computer infrastructure, to better capture attacks and defences that target the computer system.

In terms of general design (Figure~\ref{fig:v1v2comparison}), the red background is made slightly lighter making it
easier on the eyes, with the symbols watermarked in the background of the card also made lighter to maintain focus on
the content of the card whilst also still highlighting the card type. Further, the language of the text in the cards
has been improved with the aim of improving clarity and reducing technical jargon that some participants had struggled
with. For example, ``attack vectors'' and ``consequences'' were replaced with ``related attacks'' and ``related defences''
respectively. Finally, defence cards are now more akin to a stop sign to aid with better distinguishing them from the attack cards.


\subsection{Relationship Clarification}

The next revision made in Version 2 of the cards is a clarification of the links between the cards to highlight
relationships between various vulnerabilities, attacks and defences, which was described to be inadequate.
Figure~\ref{fig:v1v2comparison} shows a comparison between Version 1 and 2 of the cards. In Version 1, identifier codes
in the top right corner were used to distinguish between types of attacks, defences and vulnerabilities \circled{1}.
The relationships between attacks, defences and vulnerabilities is clarified by linking attack and defence codes within
the vulnerability cards \circled{2}. In Version 2, card identifiers are replaced with a symbol and card number related
to each type positioned in the top left corner \circled{4}. The card number is the number of the card within a type. For
example, {\em Command / Data Injection} is card number 2 for the {\em Injection} attack type. The links between
vulnerabilities follow a similar approach to Version 1, but with the codes replaced by pairs of symbols with ID numbers and
related vulnerabilities are detailed within attack and defence cards \circled{3}.
Links to vulnerabilities are given to encourage users to use intuition to make links by learning from card content, rather
than looking at explicit relationships.

\section{Evaluation -- Version 2}
\label{sec:evaluation2}

\begin{figure}[h]
    \centering
    \includegraphics[width=0.65\linewidth]{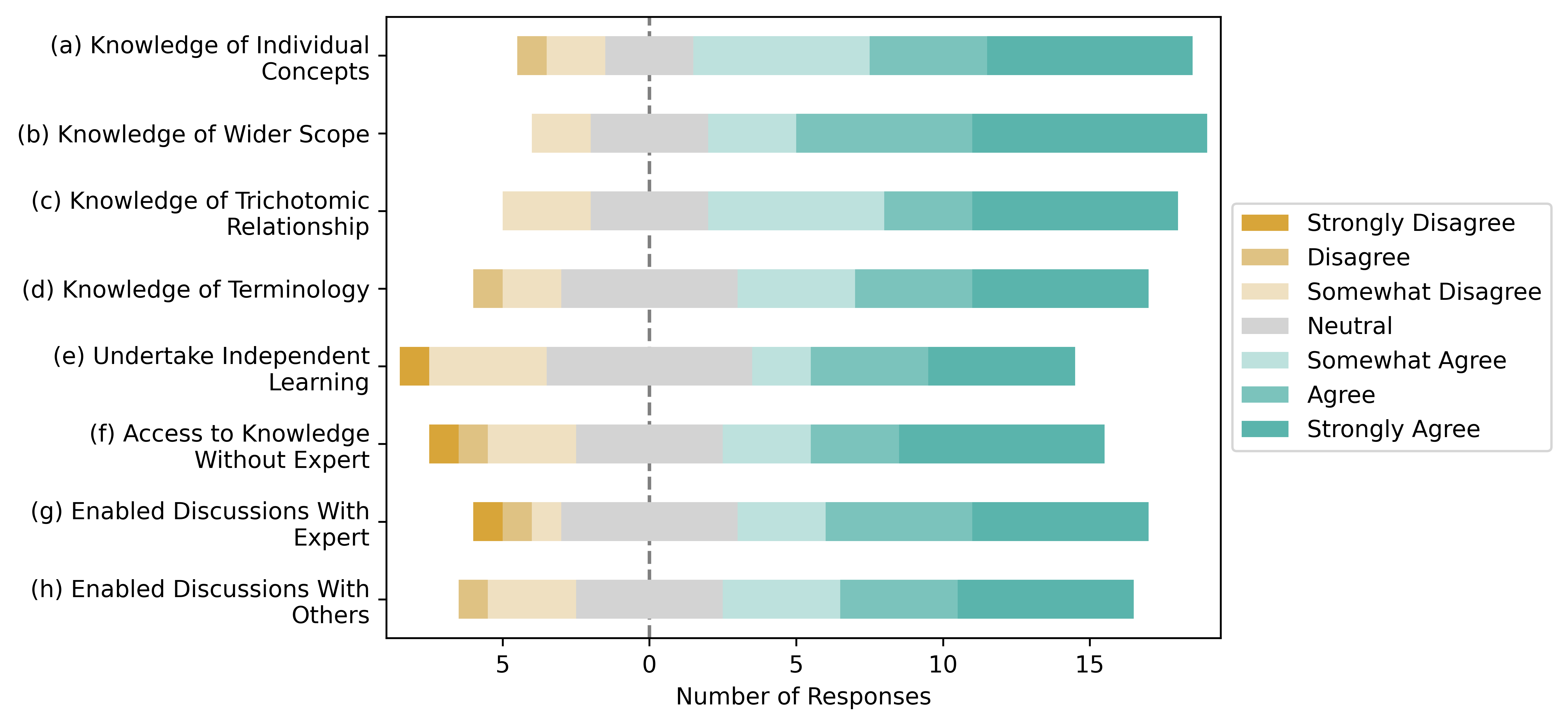}
    \caption{\centering Results of Workshop 2. This figure shows an overview of the responses for Q1 in Appendix~\ref{app:questionnaire}.}
    \label{fig:ws2likert}
\end{figure}

To evaluate the second version of the cards, we make use of the same workshop format and methodology (Table~\ref{table:workshopstructure}) as described in
Section~\ref{sec:evaluation}. In this second workshop, we recruited 23 participants with ages ranging from 10 to 15 years old (mean 12.8 years) from
either senior primary school or early high school. When asked to describe any experience and/or
skills they may already have with cybersecurity, one participant stated they know the basics, one stated that
passwords are important, one stated they know about hacking, and all other participants saying they have either no
experience (17) or are unsure (3). When asked about experience or skills with coding, two participants stated they had
limited experience with Python programming, six stating they have used Scratch or other block-based visual programming
tools, with the remainder (15) having no programming experience. Only 2 participants stated they had
experience with secure coding, with one neutral and the remainder either disagreeing or strongly disagreeing with this.
While the demographic of participants in this second workshop differs from the first workshop,
with participants in this workshop being younger, their experience with cybersecurity and code security remains similar and the activity within
both workshops are the same following the theme of {\em Code Security}. After the workshop, the participants were given the same questionnaire to answer as those received in the first workshop
(Appendix~\ref{app:questionnaire}). For subsequent discussion, we will discuss the results of the evaluation on the
second version of the cards on the same themes as the first version presented in Section~\ref{sec:results1}. The data
from questionnaires is presented as italics in quotation marks, alongside a participant ID (e.g. W2P6 for this second
workshop) where relevant. An overview of the results for the second evaluation can be seen in Figure~\ref{fig:ws2likert}.

\subsection{Results}
\label{sec:results2}

\subsubsection{Providing a Knowledge Base}
\label{sec:provideknowledge2}

For the first theme of determining whether the cards provided introductory cybersecurity knowledge that supports
learning in a well-documented manner ($RQ1$), we look at the four items (\textbf{a--d}) in Question 1 of the questionnaire in
Appendix~\ref{app:questionnaire}. For individual concepts \textbf{(a)}, 17 of the participants agreed that the cards
provided them with knowledge of cybersecurity concepts, terminology and topics (6 somewhat agree, 4 agree and 7 strongly
agree) with 3 participants scoring neutral and 3 somewhat disagreeing with this. For wider scope \textbf{(b)}, 17
participants agreed (3 somewhat agree, 6 agree and 8 strongly agree) that the cards provided them with this knowledge,
with 2 disagreeing and 4 remaining neutral. For relationships between attacks, defences and vulnerabilities \textbf{(c)},
16 agreed (6 somewhat agree, 3 agree and 7 strongly agree) that the cards facilitated the understanding of the links
between cards that capture the vulnerability-attack-defence trichotomy that is present in cybersecurity ($RQ4$), with 4 neutral
and 3 disagreeing. Finally, with regarding to cybersecurity terminology \textbf{(d)}, 14 participants agreed (4 somewhat
agree, 4 agree and 6 strongly agree) that the cards provided them with knowledge on terminology. Three participants
disagreed with this and the remaining 6 were neutral.


\subsubsection{Independent Learning and Self-Efficacy}
\label{sec:independentlearning2}

The theme of independent learning and self-efficacy relate to items (\textbf{e,f}) in Question 1. The first item
asked participants if the cards helped them undertake independent learning of cybersecurity \textbf{(e)}. We found that
for the second version of the cards, 11 participants agreed (2 somewhat agree, 4 agree, 5 strongly agree), 7 were
neutral and 5 disagreeing (1 strongly disagreeing). For the second item, we asked if the cards provided access to
cybersecurity knowledge without the presence of an expert. We found that 13 participants agreed (3 somewhat agree, 3
agree and 7 strongly agree), with 5 participants neutral to this and 5 disagreeing (1 strongly disagreeing).
These results show that the second version of our cards successfully supports independent learning and self-efficacy ($RQ2$).


\subsubsection{Providing an Interface for Discussion}
\label{sec:provideinterface2}

The final theme explored relates to the cards providing an interface for discussing key cybersecurity topics ($RQ3$) and relates
to items (\textbf{g, h}) in Question 1. The first question related to this theme asked participants if the cards enabled
them to discuss cybersecurity topics with an expert \textbf{g}. We found that 14 participants agreed with this (3
somewhat agree, 5 agree and 6 strongly agree), with 6 participants remaining neutral to this and 3 disagreeing (1
strongly disagreeing). The final question in this theme asked them if the cards enabled them to hold discussions on key
cybersecurity topics with others in their group \textbf{h}. We found that 14 participants agree with this (4 somewhat
agree, 4 agree and 6 strongly agree), with 5 participants neutral and 4 disagreeing with this.


\subsubsection{Understanding Drawbacks of Version 2}
\label{sec:limitations2}

As done in the first workshop, the next three questions were designed to gather an understanding of any
limitations to the design of the cards and whether there may be any improvements that can be made.
In Question 2, we asked participants whether there were any categories or subset
of cards not used by them in the workshop. The difference in this question compared to the same questionnaire employed
for evaluating Version 1 is due to the removal of the General cards and that they are replaced with a Glossary. Thus, in
Appendix~\ref{app:questionnaire}, we refer to {\em Question 2 (Version 2 Evaluation)} which includes the Glossary as an
option, as well as cards within a specific category for each of the classes. We found that the least used items were the
Glossary (7 participants) and Race Condition (7) cards, with the next least used cards being Injection (6), Memory (6),
Mitigation (6) and Education (6). Looking at Question 3 to understand why these types of cards (excluding the glossary)
in particular were not used, some participants mentioned that they {\em ``didn't need them''} (W2P8), were
{\em ``unnecessary''} (W2P20) or {\em ``don't know''} (W2P9). This was because some participants had decided to focus on
a different aspect of cybersecurity.
With regard to the glossary, some participants stated that {\em ``nobody reads
the glossary''} (W2P10), suggesting that this was not needed and the cards alone for some participants were enough to
facilitate learning and understanding of key cybersecurity topics. Interestingly, some participants had stated that
they had {\em ``used all the cards''} (W2P4, W2P6), with some others giving a similar reason yet selecting many (if not all)
checkboxes for all cards. In Question 4, we asked participants if there were any other improvements to the cards they
would like to suggest. Most participants (14) said {\em ``No''} to this question, however one participant suggested to use
{\em ``simple wording''}, which may correlate with why certain cards (e.g. Race Condition or Memory attacks) were not used.
As well as this, three participants suggested to make the cards {\em ``easier to use''} (W2P2, W2P14), however any suggestions as
to how they might think this could be achieved were not elaborated.



\section{Discussion}
\label{sec:discussion}

The aim of this research is to understand how a playing cards approach can provide clear communication of key
cybersecurity topics and an interface for discussing them, leveraging the information from the CyBOK knowledge base. To
investigate this, we were guided by the research questions proposed in Section~\ref{sec:cards}.
Subsequent discussion will encompass the evaluations of both versions of the cards and follows
each of the research questions in turn.

\subsection*{\normalsize RQ1: Do the cybersecurity cards provide introductory cybersecurity knowledge to novice users?}

Cybersecurity is often overlooked as a subject due to issues such as perceived technical difficulty, steep learning
curves and a requirement of specialist knowledge and/or expertise. In industry, for example, the lack of cybersecurity
professionals has been linked to a lack of practical cybersecurity content within learning
materials~\cite{caldwell2013plugging}. While CyBOK aims to rectify this learning gap, traversing the knowledge base and
understanding the material requires prior knowledge, as evidenced by the primary usage of CyBOK in the development of
higher education programmes~\cite{hallett2018mirror}. Ultimately, this means it may not be considered accessible for
novice individuals. In this work, we show that our cybersecurity cards achieved a
positive result with regard to providing introductory knowledge of key cybersecurity topics to novice users. Specifically,
we found that 82\% of participants from the first workshop in higher education agreed with this (Section~\ref{sec:provideknowledge1}).
In our second workshop involving late-primary and secondary aged school children. We found that around 70\% of them also
agreed with this (Section~\ref{sec:provideknowledge2}). Both of these results are significant as participants described
themselves initially as having little-to-no cybersecurity experience.

\subsection*{\normalsize RQ2: Do the cybersecurity cards provide material for expressing interpretation of key topics that supports independent learning and self-efficacy?}

One of the concerns surrounding cybersecurity relates to the preconception of steep learning curves and a requirement of
specialist knowledge and expertise~\cite{asen2019you,bahizad2020risks}, which is a problem that is not adequately
managed by the CyBOK knowledge base. From our evaluation of the cards, we found that most of our participants in higher
education from the first workshop agreed they were able to understand key cybersecurity topics independently (Section~\ref{sec:independentlearning1}).
In the second workshop, we found that half of the primary and secondary participants also agreed with this
(Section~\ref{sec:independentlearning2}).
It appears that removing the general cards and replacing them with a
glossary may have impacted this, backed up by statements from the participants saying that nobody
reads the glossary.
In terms of
self-efficacy, we found that participants in the first workshop agreed that even without a cybersecurity expert present
in the group, they were able to access cybersecurity knowledge solely using our cybersecurity cards. In the second
workshop, we found that more participants agreed on this compared to understanding topics independently.

\subsection*{\normalsize RQ3: Do the cards act as an index for the CyBOK knowledge base, which provides an interface for discussion on key cybersecurity topics?}

CyBOK has been shown to lack depth of cybersecurity knowledge and is hard to traverse, particularly in the aspect
of practical experience such as discussing key topics, which is essential for mastering cybersecurity
skills~\cite{manson2014case,hallett2018mirror}. The use of playing card formats as an alternative approach for learning cybersecurity topics has
shown to be successful, but are either designed for those with existing cybersecurity
knowledge~\cite{swann2021open,trickel2017shell,cybokctf} or do not leverage a peer-reviewed and well-established
knowledge foundation such as CyBOK~\cite{denning2013control,anvik2019program}. This is important as the value,
actionability and perception of security information strongly depends on the
source~\cite{rader2015identifying,redmiles2020comprehensive}. While in this work we also propose the use of a playing
cards format, we leverage a strong information foundation and the intended use is not in the context of games which
other playing cards approaches are. Furthermore, our cards have a single target persona of someone with little-to-no
expertise in cybersecurity that writes software code.
The first version of our cybersecurity cards show that the majority of participants agreed they were able to discuss
key topics with cybersecurity experts that periodically checked in on them, as well as with other members in their
group (Section~\ref{sec:provideinterface1}). In Version 2, we found similar results with more strongly agreeing compared
to those in higher education using the previous cards. While in both cases some participants disagreed with this,
this may be due to reasons such as simply not wanting to discuss topics with the experts. Interestingly, in the second
workshop, we found that more participants disagreed that the cards enabled them to hold discussions with others in their
group. This could be due to a lack of interest in doing so, or potentially they simply did not know they could do that.
In this work, we demonstrate that our playing cards approach can act as a suitable index for CyBOK and provides an
interface for discussing key cybersecurity topics for even novice or non-technical users.

\subsection*{\normalsize RQ4: Do the cards provide links between key cybersecurity topics, allowing for the capture of various scenarios?}

In the evaluation questionnaire, we asked participants whether the cards provided them with knowledge about
the relationships between attacks, defences and vulnerabilities (Section~\ref{sec:provideknowledge1}). We
found that the majority of participants agreed with this. This is further strengthened given that most
participants in the first workshop also agreed the cards promoted discussion of key cybersecurity topics with both
experts and other members in their groups (Section~\ref{sec:provideinterface1}), with similar results seen in the
second workshop (Section~\ref{sec:provideinterface2}). While the
CyBOK knowledge base does encapsulate the attack-defence-vulnerability trichotomy, it is difficult to identify
whether some topics focus on a single predominant theme or whether it spans across various
themes~\cite{gonzalez2022exploring}. Furthermore, it has been shown that because of this difficulty, these links could
be identified from series of keywords which are only meaningfully extracted via specialised algorithms such as topic
model analysis~\cite{gonzalez2022exploring,hallett2018mirror}. With regard to other learning approaches for
cybersecurity, Capture The Flag (CTF) activities can aid with highlighting links between attacks, defences and
vulnerabilities by prioritising the focus of finding vulnerabilities~\cite{trickel2017shell,swann2021open,cybokctf}.
However, the disadvantage to CTF approaches relates to requiring technical expertise (e.g. using command-line tools)
in order to progress in finding vulnerabilities~\cite{mcdaniel2016capture}. In this work, we found that
those with little-to-no cybersecurity or coding expertise can use our cards to
make links across the cybersecurity trichotomy, whilst devising 
various cybersecurity scenarios in the domain of code security.



\subsection{Reported Limitations}

From the evaluation of the first version of the cards, we identified some limitations from participant responses. First,
some participants felt the overall number of cards was too high (124 cards in total) and General cards were not used.
In the second version, the deck size was reduced to improve physical handling, whilst also adhering to
the primary goals, with General cards also replaced by a glossary. In the
second workshop, we found that there were no further suggestions to reduce the size of the deck.
Second, the layout and content of first version of the cards were described as difficult to go through by
participants in the first workshop. In version 2, we improved the contrast of the red colour,
as well as the layout of card elements to help improve clarity, as well as improving the
terminology used to describe each of the cards. In the second workshop,
none of the participants suggested any further design changes.
Finally, we believed the issue of readability in Version 1 to form links from
vulnerabilities was likely due to the identifier codes and instead replaced them with a symbol and index number.

\subsection{Future Work}

A first point of future work would be to revisit the glossary, as some participants indicated that the glossary was not
useful and in some cases not used at all. Given that the purpose of the glossary is to improve meaning and uniformity
in the usage of technical terminology, one potential solution to improve our approach is to improve the content
of the glossary (e.g. the language). As well as this, the importance of the glossary can also be better highlighted
when using the cards. While the current approach makes links to the glossary using the class symbol (e.g. attack) may
be suitable, better engagement with the glossary could be achieved through informing
users of the importance of the glossary when introducing the cards. In the case of digital cards, hyperlinks within cards
(e.g. via the symbol) to the glossary could be made clear for improved accessiblity. With regard to the layout and
written content, the language used for each of the cards can be further refined. As well as this, another design change
can be to better visibly distinguish the classes of cards. In previous work, it has been studied whether different
colours of warnings can affect one's perception of risk~\cite{leonard1999does,arenas2014color}. For example, the same
red color could be used for attack cards as it is typically associated with danger and a higher perceived relative
amount of risk. An orange or amber colour is typically used for warnings and the colour yellow for caution, which could
be used to signal the vulnerability cards.
The defence cards could be represented with a blue colour. For example, blue has represented defence in some areas of
military, such as to signal a friendly icon for NATO APP-6/A affiliation~\cite{varga2019exploration}.

\section{Conclusion}
\label{sec:conclusion}

Cybersecurity is a complex subject area that is constantly changing due to the
dynamic nature of vulnerabilities, attacks and defences. Many users who may have little-to-no knowledge
of cybersecurity are left vulnerable, with the key question of whether best practices are truly understood
remains unclear. Existing knowledge bases such as CyBOK provide key cybersecurity information, but are typically designed
to support the development of educational curricula or those with prior knowledge.
In this work, we propose an approach leveraging a playing cards format with the goals of introducing
cybersecurity topics to novice users, facilitating independent learning and understanding of
the various relationships found in the cybersecurity ecosystem. Upon evaluation, we found that our
approach was successful in achieving these goals for a wide age group (10-35 years) of both non-technical
users and those with some experience. Using the data from this evaluation,
we designed a second version of these cards and further evaluated them and showing they still meet the proposed
requirements. Ultimately, our cybersecurity cards provide a comprehensive and effective tool that
allows novice individuals to gain introductory knowledge of cybersecurity, while promoting understanding and independent
learning of cybersecurity.




\bibliographystyle{ACM-Reference-Format}
\bibliography{references}

\appendix



\section{Workshop Evaluation Questionnaire}
\label{app:questionnaire}

\paragraphb{Question 1}\par\noindent
Each item below is scored using a Likert-type scale (1 - 7): (1=Strongly disagree, 2= disagree, 3= somewhat disagree, 4= don't know/neutral, 5= somewhat agree, 6= agree, 7= strongly agree).
\par\vspace{0.5cm}\par\noindent
The Cybersecurity cards that I have used during the workshop:

\begin{longtable}{ p{\mwide{}} >{\centering\arraybackslash}p{\thin{}} >{\centering\arraybackslash}p{\thin{}} >{\centering\arraybackslash}p{\thin{}} >{\centering\arraybackslash}p{\thin{}} >{\centering\arraybackslash}p{\thin{}} >{\centering\arraybackslash}p{\thin{}} >{\centering\arraybackslash}p{\thin{}} }
    & Strongly Disagree & Disagree & Somewhat Disagree & Don't Know / Neutral & Somewhat Agree & Agree & Strongly Agree \\
    \st{\textbf{(a)} Provided me with knowledge about individual cybersecurity concepts }\\
    \st{\textbf{(b)} Provided me with knowledge about the wide scope of cybersecurity concepts }\\
    \st{\textbf{(c)} Provided me with knowledge about the relationship between cybersecurity concepts, such as the relationships between attacks, defences and vulnerabilities }\\
    \st{\textbf{(d)} Provided me with knowledge about cybersecurity terminology }\\
    \st{\textbf{(e)} Enabled me to undertake independent learning about cybersecurity }\\
    \st{\textbf{(f)} Provided access to cybersecurity knowledge when the cybersecurity expert was not present }\\
    \st{\textbf{(g)} Enabled me to discuss cybersecurity with the cybersecurity expert } \\
    \st{\textbf{(h)} Enabled me to discuss cybersecurity with others (i.e. not including the cybersecurity expert }\\
\end{longtable}

\par\paragraphb{Question 2}\par\noindent
Which category or subset of the cybersecurity cards did you not use, if any? Please tick all the boxes that apply in your opinion:
\begin{enumerate}[label=$\square$]
    \item General Attack Cards
    \item General Vulnerability Cards
    \item General Defence Target Cards
    \item General Defence Cards
    \item Detailed Attack Cards
    \item Detailed Vulnerability Cards
    \item Detailed Defence Cards
\end{enumerate}

\par\paragraphb{Question 2 (Version 2 Evaluation)}\par\noindent
Which category or subset of the cybersecurity cards did you not use, if any? Please tick all the boxes that apply in your opinion:
\par\noindent{\em The logos in the Glossary indicate the specific categories of Attack, Defence and Vulnerability cards.}
\begin{enumerate}[label=$\square$]
    \item Glossary
    \item Attack cards (in general) (fully red cards)
    \item Vulnerability cards (in general) (cards with half red/half white diagonal)
    \item Defence cards (in general) (white cards with red border)
    \item Attack -- Injection cards
    \item Attack -- Memory cards
    \item Attack -- Race condition cards
    \item Attack -- Side channel cards
    \item Attack -- Authentication cards
    \item Attack -- Web cards
    \item Attack -- System cards
    \item Attack -- Human factors cards
    \item Defence -- Detection cards
    \item Defence -- Mitigation cards
    \item Defence -- Education cards
    \item Defence -- Prevention cards
    \item Vulnerability -- Code cards
    \item Vulnerability -- System cards
    \item Vulnerability -- Environment cards
    \item Vulnerability -- User cards
    \item Vulnerability -- Management cards
    \item \noindent\fbox{\begin{minipage}{\linewidth} Other\par\vspace{0.25cm}\end{minipage}}
\end{enumerate}

\clearpage

\par\paragraphb{Question 3}\par\noindent
Looking at the types of cards you selected in the previous question, why did you \textbf{not} use those cards?\par\vspace{0.5cm}

\noindent\fbox{\begin{minipage}{\linewidth}
                   Enter your answer\par\vspace{0.25cm}
\end{minipage}}

\vspace{0.5cm}

\par\paragraphb{Question 4}\par\noindent
How could the \textbf{cybersecurity cards potentially be improved} in your opinion? Please tick all the boxes of the statements
below you agree with:
\begin{enumerate}[label=$\square$]
    \item The total number of cards was too high.
    \item The total number of cards was too low.
    \item There were too many different types/categories of cybersecurity cards.
    \item The logos and icons were not clear.
    \item The numbering of the cards was not clear.
    \item The colour-coding of the different types/categories of cybersecurity cards was not clear.
    \item The relationships between the different cybersecurity cards was not clear.
    \item The information provided by the cybersecurity cards was too abstract.
    \item The information provided by the cybersecurity cards was too detailed.
    \item The information provided by the cybersecurity cards was too technical or too difficult to understand.
    \item {\em The Glossary overview of the different types of cybersecurity cards was not clear. (Version 2)}
    \item \item \noindent\fbox{\begin{minipage}{\linewidth} Other\par\vspace{0.25cm}\end{minipage}}
\end{enumerate}

\par\paragraphb{Question 5}\par\noindent
Are there any other improvements to cybersecurity cards you would like to suggest?\par\vspace{0.5cm}

\noindent\fbox{\begin{minipage}{\linewidth}
                   Enter your answer\par\vspace{0.25cm}
\end{minipage}}

\section{Version 2 Glossary}
\label{app:glossary}

\includegraphics[scale=0.75,angle=-90]{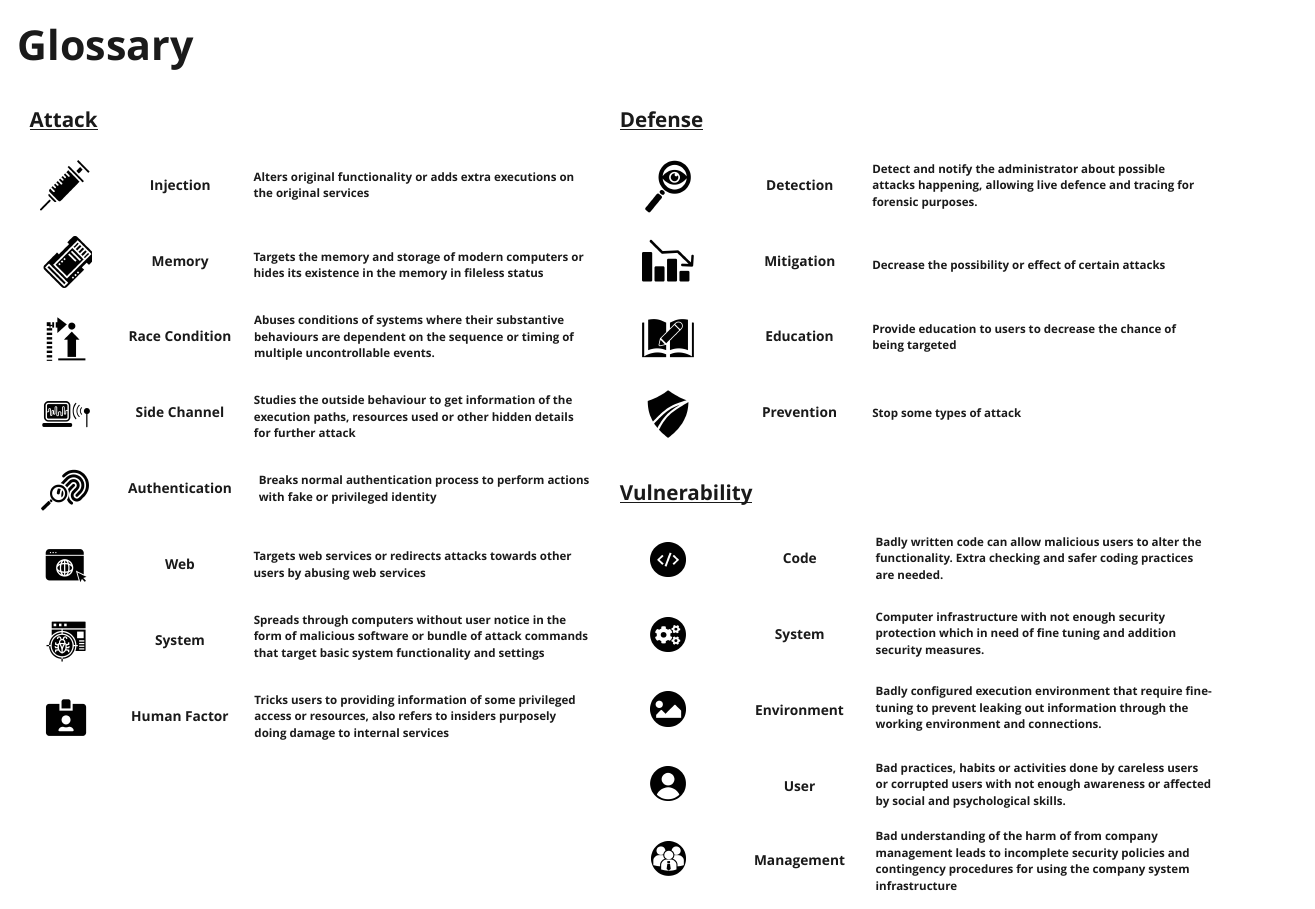}

\end{document}